\documentclass[twocolumn, aps,preprintnumbers,amsmath,amssymb,nofootinbib,prl]{revtex4-1}

\usepackage{graphicx}
\usepackage{epsfig}
\usepackage{textcomp}

\def\lsim{\mathrel{\rlap{\lower4pt\hbox{\hskip1pt$\sim$}}
    \raise1pt\hbox{$<$}}}         
\def\gsim{\mathrel{\rlap{\lower4pt\hbox{\hskip1pt$\sim$}}
    \raise1pt\hbox{$>$}}}         

\begin{document}

\preprint{\today}

\title{Critical comment on the recent microscopic black hole search at the LHC}


\author{Seong Chan Park}
\email{sc@jnu.ac.kr}
\vspace{1cm}
\affiliation{Department of Physics, Chonnam National University, 300 Yongbong-dong, Buk-gu, Gwangju, 500-757, Korea
}

\vspace{1.0cm}

\begin{abstract}
Recently the CMS collaboration at the LHC reported ``the first direct limit on black hole production at a particle accelerator" using a data sample corresponding to an integrated luminosity of $35 {\rm pb}^{-1}$ of $pp$ collision at a center of mass energy of $7$ TeV \cite{CMS}. Even though the result has a strong impact on future searches, the interpretation lacks enough theoretical support. In this letter, we show that the parameter range which was considered  by the CMS collaboration is actually out of the validity range of semi-classical black hole picture so that the Monte-Carlo simulation result which was crucially used in the analysis still needs further solid scientific basis.  
\end{abstract}


\maketitle

\newpage

\section{Introduction} 
Recently the CMS collaboration reported the first direct limits on the  microscopic black hole masses using a  data sample corresponding to an integrated luminosity of 35 ${\rm pb}^{-1}$ at a center-of-mass energy of 7 TeV of $pp$ collisions at the LHC \cite{CMS}. 
Events with the large total scalar sum of the transverse energy, $S_T$, of isolated high energy jets, leptons (electrons and muons), and photons were analyzed, which are  typically expected from the decay of a semi-classical black hole.
The reported limits on the microscopic black hole mass are in the range around 3.5-4.5 TeV in a model with large extra dimensions.  Indeed low energy gravity scenarios based on large or warped extra dimension(s) \cite{ADD,RS} predict that a high energy collider, such as the LHC, may have a chance  to produce a sizable number of microscopic black holes \cite{Giddings,Dimopoulos} if  the collision energy would be  exceedingly larger than the scale where higher dimensional gravity becomes strong \cite{'tHooft}.  The result has a strong impact on future searches for low energy quantum gravity and extra dimensions. Considering the huge significance, we carefully re-examine the theoretical background behind the search and show how the current result is (not) supported by existing theoretical studies and scientific reasonings.

\section{Critical review on theoretical background}
The precise prediction regarding the microscopic black hole at colliders is not currently available as it depends on unknown details of geometry of extra dimension and quantum nature of gravity. However, still some generic features  are understood  by semi-classical approximation of higher dimensional rotating black hole~\cite{Myers-Perry}. The semi-classical approximation is valid provided that the size of event horizon, $r_H$, is significantly larger than the typical distance scale of the relevant gravity theory ($\ell_G \sim 1/M_G$), in which case the quantum gravitational corrections are expected to be small and negligible: $\delta_{QG}\sim ( \ell_G/r_H)^p \ll 1$ with a positive power $p>0$. 
In particular, if the size of horizon is much smaller than the size of extra dimensions or compactification radius, $ r_c$, higher dimensional physics is relevant and  the scale $M_G$ should be identified with the scale of $D$-dimensional gravity ($M_D \sim 1/\ell_D$). In short,  validity of semi-classical, higher dimensional black hole picture is justified for a black hole of the mass $M$ satisfying the following condition:
\begin{eqnarray}
\ell_D \ll r_{BH}(M) \ll r_c.
\label{eq:cons}
\end{eqnarray}
The former condition (i.e., $r_{BH}\gg 1/M_D$) ensures that the object under consideration can be treated as a semi-classical object and the black hole solution to general relativity provides a good approximation. The later condition (i.e., $r_{BH} \ll r_c$) ensures that the black hole is actually a higher dimensional object. More explicitly, the above conditions are satisfied when the mass of black hole ($M$) is in the following range:
\begin{eqnarray}
1\ll \frac{M}{M_D}  \ll \left(\frac{M_{\rm Planck}}{M_D}\right)^{2 \left(\tfrac{D-3}{D-4}\right)},\label{eq:condition}
\end{eqnarray}
 where we used the relation $M_{\rm Planck}^2 \sim M_D^{D-2}r_c^{D-4}$ with the Planck scale ($M_{\rm Planck}\sim 10^{16}$ TeV) for the compactification radius $r_c$. The size of black hole is roughly given by the Schwarzschild radius $r_{BH}\simeq r_{Sch}= C_D \left(M/M_D\right)^{1/(D-3)}\ell_D$ where $C_D$ is a numerical constant of the order of unity\footnote{Explicitly, $C(D)=\frac{1}{\sqrt{\pi}} \left(\frac{8\Gamma((D-1)/2)}{D-2}\right)^{1/(D-3)}\in (0.75, 0.92)$ for $D\in [6,10]$.} for a  black hole with a given mass $M$ in $D$-dimensions. As the Planck scale is hierarchically greater than the scale of higher dimensional gravity in low energy gravity models, there exists a large parameter window for   $M$. For instance, if $M_D \sim 1$ TeV, as for the solution to the hierarchy problem in extra dimension models, the validity range of classical higher dimensional blackhole picture is huge as $ 10^{32\left(\tfrac{D-3}{D-4}\right)}\gg M [{\rm TeV}]\gg 1 $.
One should note that the limit in Eq. \eqref{eq:condition} is consistent with the assumption that the action for higher dimensional gravity is well described by the leading order Einstein-Hilbert action and the higher order curvature terms are regarded as small corrections, $\Delta \ll 1$:
\begin{eqnarray}
S_G =\int d^D x \sqrt{-g} \frac{M_D^{D-2}}{2} R \left[1 + \Delta \right], 
\label{eq:action}
\end{eqnarray}
where  the higher oder correction term, $\Delta$,  is expanded as $\Delta = C_1 \frac{R}{M_D^2} +C_2 \frac{R^2}{M_D^4}  \cdots$, with dimensionless coefficients $C_{n\geq 1}$ which are supposed to be all ${\cal O}(1)$ \footnote{The higher order terms can be also regarded as  higher derivative terms as the curvature is second derivatives of the metric $R\sim \partial \partial g$.}. We can see that perturbativity is guaranteed for a `large' black hole with a large mass ($M\gg M_D$) as the corresponding curvature at the horizon is small for a large black hole :
\begin{eqnarray}
\Delta\sim \frac{R}{M_D^2} \sim \frac{1}{r_{BH}^2 M_D^2}\sim \left(\frac{M_D}{M}\right)^{\tfrac{2}{D-3}} \ll 1.
\label{eq:cons3}
\end{eqnarray}

When we consider black holes which are formed by particle collisions in Trans-Planckian energy range ($\sqrt{s}\simeq M \gg M_D$), its mass  is essentially determined by collision energy \footnote{Some large fraction of energy is expected to be radiated away in the form of gravitational wave too.} and the production cross section is well approximated by a  geometric cross section $\sigma \sim \pi r_{BH}(\sqrt{s})^2$. The validity of geometric approximation was originally conjectured by Kip Thorne \cite{hoop} and more precisely proved later \cite{formation1, formation2, formation3} in semi-classical domain \footnote{Actually the production cross section can be obtained taking angular momentum of black hole into account \cite{Park2001, IOP1}.}.

After the initial stage of formation, black hole loses its energy and angular momentum through Hawking radiation \cite{Hawking} mainly to the standard model particles \cite{Emparan}.   The detailed decay rates, which are essential in analyzing black hole signatures at the LHC, crucially depend on greybody factors, ${}_s\Gamma_{\ell m}$, and thermal factor with angular velocity at the horizon $\Omega$, as: 
\begin{eqnarray}
\frac{dE_{s,\ell,m}}{dt d\omega} =\frac{1}{2\pi}\frac{{}_s\Gamma_{\ell m}}{e^{(\omega-m \Omega)/T}-(-1)^{2s}}
\end{eqnarray} 
where $\omega, s,\ell,m$ stands for the energy, spin and angular quantum numbers of radiated particle, respectively. In semi-classical domain,  greybody factors have been obtained for an arbitrary spin ($s=0,1/2,1$ and also $s=2$ in part) in \cite{IOP1,IOP2,IOP3} and \cite{Kanti1,Kanti2,Kanti3} \footnote{Gravitational radiation through Hawking radiation is still missing. There are some recent developments \cite{graviton1, graviton2, graviton3}}.
 Naively the radiation is considered to be democratic (i.e. same rate to each species) but the actual spectrum significantly depends on the detailed status of black hole, for instance its angular momentum.  In obtaining greybody factors, it is assumed that the back-reaction from the Hawking radiation is negligibly small. This assumption is justified only for a large black hole with $M\gg M_D$.

The prominent  thermodynamic features of semi-classical microscopic black hole is encapsulated in its temperature ($T \sim 1/r_{BH}\sim {\cal O}(100) {\rm GeV}$) and its large entropy or equivalently horizon area ($S \sim (M_D r_{BH})^{D-2} \gg 1$). The large entropy ensures that the consistency of the curvature perturbation since  the correction term is related to the entropy as
%
$\Delta \sim 1/S^{2/(D-2)}$.
%
The large entropy is a direct consequence of  the required consistency condition in Eq. \eqref{eq:cons}  and leads the large multiplicities of particles in decay processes. Actually the large multiplicity is one of the most important selecting criteria for black hole search \cite{CMS}. With all these semi-classical understanding of microscopic black hole, semi-realistic Monte-Carlo (MC) event generators for black holes have been developed and used by the CMS collaboration as well as ATLAS collaboration. Two of the most developed MC generators, called BlackMax \cite{blackmax} and CHARYBDIS \cite{CHARYBDIS, CHARYBDIS2}, are actually used in the recent CMS search for microscopic black holes. We emphasize that these MC event generators are applicable  only for  large black holes.

\section{Validity of semi-classical approximation in the CMS result}

Now we are ready to comment on the recent CMS result more explicitly based on the discussions so far. Our emphasis is on  the validity of semi-classical approximation and the applicability of the MC event generators.

First of all,  we would point out that the parameter range in the CMS result  is actually out of the validity range of semi-classical approximation of black hole.   As is found in Table I in Ref. \cite{CMS},  three values of `fundamental scale' are considered :$M_D^{\rm CMS}=1.5, 2.0, 3.0$ TeV.  The corresponding threshhold  masses for the black hole formations are taken $M_{BH}^{\rm CMS}= 2.5-5.0, 2.5-5.0, 3.5-5.0$ TeV for various dimensions $D=6-10$, respectively. The ratio of the minimum black hole mass and the fundamental scale which was excluded by the CMS data is in the range \cite{CMS}
\begin{eqnarray}
\left(\frac{M}{M_D}\right)_{\rm CMS} &\in& \left[\frac{3.5}{3.0}, \frac{4.0}{1.5}\right]_{D=6},\\
&\in&\left[\frac{4.0}{3.5}, \frac{4.5}{1.5}\right]_{D=10},
\label{eq:CMS}
\end{eqnarray}
%
or the corresponding size of Schwarzschild radius is
\begin{eqnarray}
\left(r_{BH}\right)_{\rm CMS} &\in& \left[0.9, 2.1\right]\ell_6,\\
&\in& \left[0.9,2.4\right]\ell_{10}.
\end{eqnarray}
%

\begin{figure}[t]
\centering
\includegraphics[width=0.49\textwidth]{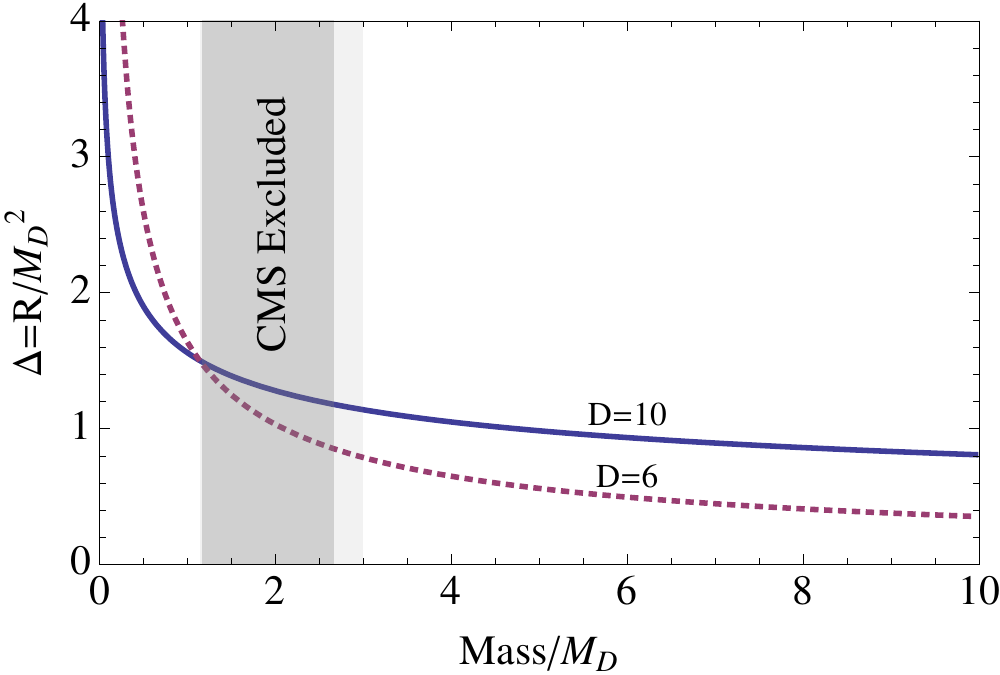}
\includegraphics[width=0.49 \textwidth]{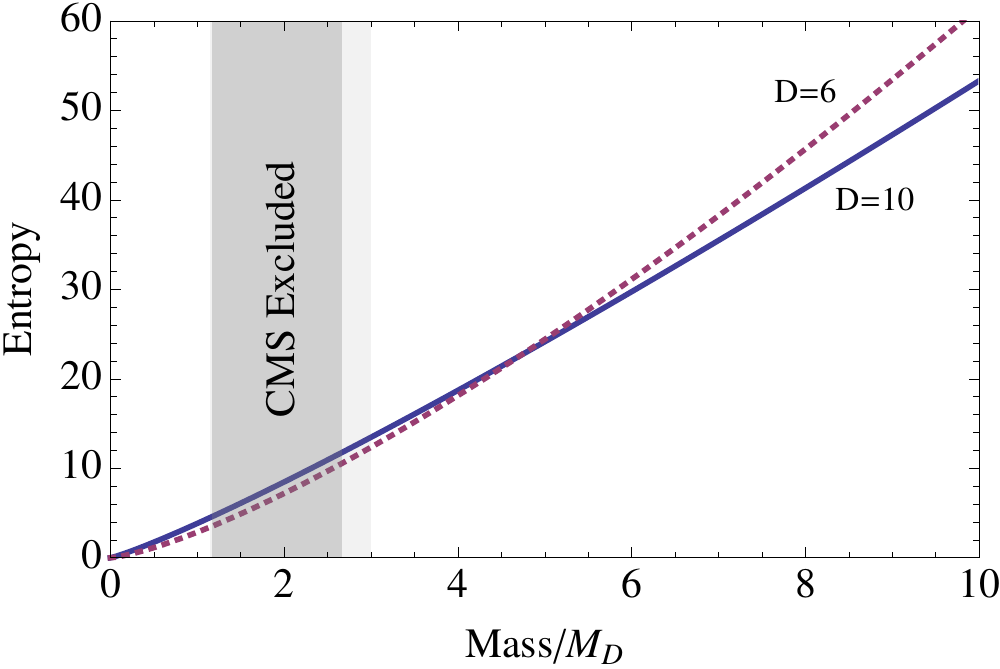}
\caption{\label{Fig:entropy} The higher order curvature term, $\Delta$, (top) and entropy(bottom) are plotted for higher dimensional black hole ($D=6$(dotted), $D=10$(solid)).  The vertical column in gray is the CMS exclusion region: $M/M_D=[3.5/3.0 (4/3.5), 4.0/1.5(4.5/1.5)]$ for $D=6(10)$, respectively.}
\label{Fig:entropy}
\end{figure}

To see the validity of semi-classical approximation in the given range, we plotted the estimation of higher order correction (upper) and extrapolated value of entropy (lower) in Fig. \ref{Fig:entropy} with respect to the given black hole masses. Note that as the mass becomes larger, the higher order correction $\Delta$ becomes smaller and the semi-classical approximation becomes more reliable. The dotted and solid lines, which respectively correspond to $D=6$ and $D=10$, behave similarly but show some deviation at the large mass region. For the consistency of the perturbative expansion, it is required that the correction term should be significantly smaller than the leading order term ($\Delta \ll 1$) and entropy should be large ($S \gg 1$). However, the CMS exclusion range (vertical columns in gray), the correction is as large as or even larger than the leading order term. Also the entropy of black hole is still less than 10 or so so that we cannot tell that the calculation is trustworthy in semi-classical sense. Within this parameter space, all the MC simulations suffer from large quantum corrections of the order of $\sim {\cal O}(M_D/M)^{p>0}$ which can lead a significant change in the final result. 

\section{Model dependence: the bulk SM in warped extra dimension}
Finally, we point out that the estimation of the black hole production cross section is highly model dependent. Here we consider a well motivated warped extra dimension model with the bulk standard model fields in \cite{Agashe:2006hk} to show the model dependence of the production cross section.  
 
\begin{figure}[t]
\centering
\includegraphics[width=0.49 \textwidth]{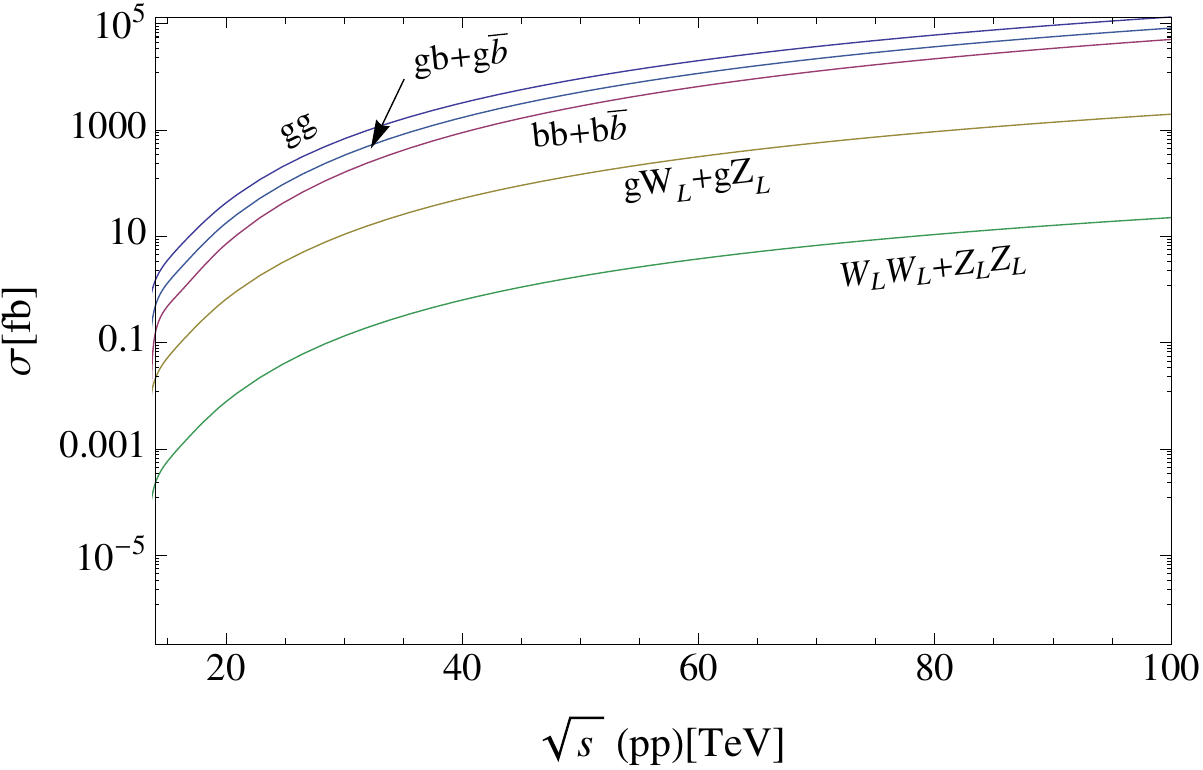}
\caption{\label{Fig:xsection} The production cross section of black hole by  $gg, gb +g\bar{b}, bb+b\bar{b}, gW_L+gZ_L, W_L W_L+Z_L Z_L$ from top to bottom in $\sqrt{s}\in (14, 100)$ TeV thinking future higher energy upgrade of the LHC.}
\end{figure}
 
 The essential feature of the model is ``light-fermion-phobic" nature of TeV-scale gravity as the wave function profiles of light fermions are all localized toward the Planck brane where gravity becomes strong at $M_{\rm Planck}$ rather than 1 TeV. As a result, the dominant chances for making black hole are  bottom quark and gluon even though gluon cross section is volume suppressed by $\sim 1/30$. The top quark contribution is PDF suppressed thus is negligible. Another important channels are by longitudinal components of weak gauge bosons following the Higgs equivalence principle. One should notice that Higgs boson (or equivalently $W_L$ and $Z_L$) should locate around the low scale brane (IR-brane) for the setup to address the hierarchy problem so that it can effectively feel TeV-gravity. However, vector boson fusion process by weak gauge bosons are all weak coupling suppressed so that they never dominate the black hole production. In Fig. \ref{Fig:xsection} we plot the cross sections for the center of mass energy in $\sqrt{s}\in (14, 100)$ TeV  by $gg, gb (g\bar{b}), bb(\bar{b}), gW_L, gZ_L, W_L W_L$ and $Z_L Z_L$ channels, respectively. For PDF convolution of parton level cross sections, we use MSTW2008 \cite{MSTW} and CTEQ \cite{CTEQ} and the results are confirmed by a modified version of  CHARYBDIS 2.0 \cite{CHARYBDIS2} \footnote{J. Frost kindly confirmed the results with CHARYBDIS2}.
 
Clearly the CMS result is not applicable to this model for two reasons. First, the estimation of the cross section is much suppressed in this model. Essentially $u\bar{u}$ and $d\bar{d}$ are the dominant production cross section in usual calculations but they are all negligible in this model .  For the low energy run with $\sqrt{s}=7$ TeV, the cross section is well below 0.1 fb.   Second, the main decay products are very distinctive in this model. Heavy quarks (top, bottom), Higgs (or equivalently longitudinal components of weak gauge bosons) and gluon but the events with light quarks are highly suppressed in this model.  

\section{conclusions}
We critically re-examined the recent microscopic black hole search  by the CMS collaboration at the LHC \cite{CMS} and found that the claimed excluded range of microscopic black hole mass  is not justified by currently available theoretical studies unfortunately. Also importantly, the estimation of the black hole production cross section and the decay pattern of produced black holes are highly model dependent. In a theoretically well motivated model-the bulk standard model in warped extra dimension-for instance, the expected production cross section is significantly below the currently detectable range at the LHC and the main decay products of black hole are multi top-quarks and Higgs particles which could not be covered by the current search strategy but requires further detailed studies \cite{Frost}.     Even though there is a chance for the LHC may be able to observe some black hole precursors  \cite{Dimopoulos:2001qe, Meade}  and low energy Kaluza-Klein excitation modes of bulk fields\footnote{An important counter example is the model in the presence of compact hyperbolic extra dimension background where  no low energy excited Kaluza-Klein modes are expected \cite{CHS}.},  we may have to wait for future higher energy run with a significantly larger collision energy than what we currently have for the feasible black hole search. 
  
{\bf Acknowledgement:} I appreciate valuable communications and discussions with K.-y. Oda, D. Ida, H. Yoshino, M. Mangano, S. Giddings, J. Frost, M. Cavaglia, D. Stojkovic, S. Mukohyama, S. Sugimoto, M. Nojiri and D. Orlando. This research was supported by Basic Science Research Program through the National Research Foundation of Korea (NRF) funded by the Ministry of Education, Science and Technology (2011-0010294). At the early stage of the work, this work was supported by the World Premier International Research Center Initiative (WPI initiative) by MEXT and also  by the Grant-in-Aid for scientific research (Young Scientists (B) 21740172) from JSPS during my stay at IPMU, the University of Tokyo.

\end{document}